\documentclass[epjST]{svjour}
\usepackage{graphics}
\begin{document}
\title{A solvable model for octupole phonons}
\author{P.~Van~Isacker\inst{1}\fnmsep\thanks{\email{isacker@ganil.fr}}}
\institute{Grand Acc\'el\'erateur National d'Ions Lourds, CEA/DRF-CNRS/IN2P3\\
Bvd Henri Becquerel, BP 55027, F-14076 Caen, France}
\abstract{
A solvable model is proposed for the description of octupole phonons in closed-shell nuclei,
formulated in terms of shell-model par\-ticle--hole excitations.
With some simple assumptions concerning single-particle energies
and two-body interactions,
closed expressions are derived for the energy and wave function of the octupole phonon.
In particular, it is shown
that the components of the octupole phonon
are proportional to Wigner $3j$ coefficients.
This analytic wave function is proven to be exactly valid in light nuclei,
which have $LS$ shell closures that coincide with those of the three-dimensional harmonic oscillator,
and to be valid to a good approximation in heavier nuclei,
which have $jj$ shell closures due to the spin--orbit interaction.
The properties of the solvable model are compared
with the results of a realistic shell-model calculation for $^{208}$Pb.}
\maketitle

\section{Introduction}
\label{s_intro}
Nuclei with a closed-shell configuration for neutrons and/or protons
frequently exhibit low-energy excitations
with angular momentum $J=3$ and negative parity.
Such excitations are associated with nuclear shapes
that break reflection symmetry
and, in particular, with pear-like or octupole shapes~\cite{Butler96}.
Given the closed-shell configuration of at least one type of nucleon,
the nucleus is thought to have a spherical equilibrium shape in its ground state
and to exhibit reflection asymmetric oscillations of the octupole type around that shape.
Nuclei with neutrons {\em and} protons in the valence shell
may acquire a permanent ground-state deformation 
and an open question is
whether they can assume a permanent pear-like deformation. 
Interest in this question was rekindled in 2013
by observed indications of such static octupole deformation
in the ground-state configuration of $^{224}$Ra~\cite{Gaffney13}.

By virtue of their supposed collective structure,
octupole excitations are thought to exhibit phonon-like behaviour~\cite{BM75},
which renders them of particular interest,
being at the cross-roads of microscopic and collective descriptions of nuclei.
Consequently, many models of nuclear octupole excitations
have been considered in the past (for a review, see Ref.~\cite{Butler96}).
From a shell-model point of view,
an octupole vibration of a closed-shell nucleus
corresponds to a coherent superposition of particle--hole excitations.
This is the basis of the description of the $3^-$ octupole state in $^{208}$Pb
proposed by Brown~\cite{Brown00},
leading to results in broad agreement with experimental findings.
Such calculations are, however, challenging when extended to more complex structures
such as multi-phonon states in terms of $n$-particle--$n$-hole excitations~\cite{Brown00}
or the coupling of particles or holes to an octupole phonon~\cite{Ralet19}.
It is therefore of interest to construct
an approximate but solvable model
of octupole phonons in terms of particle--hole excitations.
This is the purpose of the present contribution,
which is structured as follows.
In Sect.~\ref{s_general} some general notions
on the description of octupole excitations in a shell-model framework
are introduced.
The solvable model for octupole excitations is described in Sect.~\ref{s_solvable},
and subsequently applied to $^{208}$Pb
and compared with the results of a realistic shell-model calculation
in Sect.~\ref{s_pb208}.
The paper concludes with a summary and outlook in Sect.~\ref{s_summ}.

\section{General considerations about octupole phonons}
\label{s_general}
It is assumed that all excitations are with reference to a doubly-closed-shell nucleus,
which is represented by $|{\rm o}\rangle$.
The hole orbitals below the shell closure
belong to a set $\{j_{\rho k},k=1,2,\dots\}$
and the set of particle orbitals above the shell closure
is $\{j_{\rho k'},k'=1,2,\dots\}$.
Particle orbitals are consistently denoted with primed indices
and hole orbitals with unprimed ones;
both occur for neutrons ($\rho=\nu$) as well as for protons ($\rho=\pi$).
A collective octupole phonon corresponds to a coherent superposition
of particle--hole excitations with respect to $|{\rm o}\rangle$,
\begin{equation}
|3^-_{\rm c}\rangle\equiv
\sum_{\rho k'k}\rho_{k'k}|j_{\rho k'}j_{\rho k}^{-1};3^-\rangle=
\sum_{\rho k'k}\rho_{k'k}[a^\dag_{ \rho k'}\times b^\dag_{\rho k}]^{(3)}|{\rm o}\rangle,
\label{e_oph}
\end{equation}
where $\rho_{k'k}$ are coefficients.
The operator $a^\dag_{j'm'}$ creates a nucleon particle in the orbital $j'$ with projection $m'$
while $b^\dag_{jm}$ creates a nucleon hole in the orbital $j$ with projection $m$.
The single-particle and single-hole states are therefore
\begin{equation}
|j'm'\rangle=a^\dag_{j'm'}|{\rm o}\rangle,
\qquad
|j^{-1}m\rangle=b^\dag_{jm}|{\rm o}\rangle,
\label{e_sphs}
\end{equation}
and particle and hole operators are related through $b^\dag_{jm}=(-)^{j+m}a_{j-m}$.

The coefficients $\rho_{k'k}$ in Eq.~(\ref{e_oph})
are expressed in the basis $|j_{\rho k'}j_{\rho k}^{-1};3^-\rangle$,
which includes neutron as well as proton particle--hole excitations.
They result from the diagonalisation of the nuclear Hamiltonian,
which generically can be written as
\begin{equation}
\hat H=
\sum_\rho\hat H_\rho+\hat V_{\nu\pi}=
\sum_\rho\left(\sum_{k'}\epsilon_{\rho k'}\hat n_{\rho k'}-\sum_k\epsilon_{\rho k}\hat n_{\rho k}+
\hat V_{\rho\rho}\right)+\hat V_{\nu\pi},
\label{e_ham}
\end{equation}
where $\epsilon_{\rho k'}$ and $\epsilon_{\rho k}$
are the single-particle energies
pertaining to the orbitals above and below the shell closure, respectively.
The structure of the octupole phonon is therefore determined by the matrix elements
\begin{equation}
\langle j_{\rho k'}j_{\rho k}^{-1};3^-|\hat H|j_{\bar\rho l'}j_{\bar\rho l}^{-1};3^-\rangle,
\label{e_me}
\end{equation}
which can be obtained by means of particle--hole conjugation.
For the matrix element with $\rho=\bar\rho$ one finds,
up to an overall diagonal constant,
\begin{eqnarray}
&&\langle j_{\rho k'}j_{\rho k}^{-1};J|\hat H_\rho|j_{\rho l'}j_{\rho l}^{-1};J\rangle
\nonumber\\&&\qquad=
\left(\epsilon_{\rho k'}-\epsilon_{\rho k}\right)\delta_{kl}\delta_{k'l'}-
\sum_R(2R+1)
\left\{\begin{array}{ccc}
j_{\rho k'}&j_{\rho k}&J\\
j_{\rho l'}&j_{\rho l}&R\end{array}\right\}
V^{\rho\rho R}_{k'll'k},
\label{e_vph1}
\end{eqnarray}
where the symbol between curly brackets is a Racah $6j$ coefficient~\cite{Shalit63,Talmi93}
and $V^{\rho\rho R}_{k'll'k}$ is a two-body matrix element of the like-nucleon interaction $\hat V_{\rho\rho}$,
\begin{equation}
V^{\rho\rho R}_{k'll'k}\equiv\langle j_{\rho k'}j_{\rho l};R|\hat V_{\rho\rho}|j_{\rho l'}j_{\rho k};R\rangle.
\label{e_vpp1}
\end{equation}
An important feature of the physics of octupole phonons is
that, due to the neutron--proton interaction,
a non-zero matrix element also exists between a neutron and a proton particle--hole excitation.
For the off-diagonal matrix element~(\ref{e_me}) with $\rho=\nu$ and $\bar\rho=\pi$ one finds
\begin{equation}
\langle j_{\nu k'}j_{\nu k}^{-1};J|\hat V_{\nu\pi}|j_{\pi l'}j_{\pi l}^{-1};J\rangle=
-\sum_R(2R+1)
\left\{\begin{array}{ccc}
j_{\nu k'}&j_{\nu k}&J\\
j_{\pi l'}&j_{\pi l}&R\end{array}\right\}
V^{\nu\pi R}_{k'll'k},
\label{e_vph2}
\end{equation}
where $V^{\nu\pi R}_{k'll'k}$ is a two-body matrix element
of the neutron--proton interaction $\hat V_{\nu\pi}$,
\begin{equation}
V^{\nu\pi R}_{k'll'k}\equiv\langle j_{\nu k'}j_{\pi l};R|\hat V_{\nu\pi}|j_{\pi l'}j_{\nu k};R\rangle=
-(-)^{j_{\pi l'}+j_{\nu k}-R}
\langle j_{\nu k'}j_{\pi l};R|\hat V_{\nu\pi}|j_{\nu k}j_{\pi l'};R\rangle.
\label{e_vpp2}
\end{equation}

The last term on the right-hand side of Eq.~(\ref{e_vph1})
and the right-hand side of Eq.~(\ref{e_vph2})
represent the Pandya transformation of the particle--particle two-body interaction~\cite{Pandya56}.
These relations can be obtained from general principles of particle--hole conjugation,
as discussed by Bell~\cite{Bell59},
and summarised in Chapter~3 of the monograph~\cite{Lawson80}.
Because of considerations of anti-symmetry,
it is worthwhile to give a careful derivation of the Pandya relation,
following the formalism of Appendix 3B of Ref.~\cite{BM69}.
Specifically, Eq.~(3B-38) of Ref.~\cite{BM69},
adapted to the present notation,
relates {\em reduced} matrix elements as follows:
\begin{eqnarray}
&&\langle j_3^{-1}j_4;J\|\hat V\|j_1^{-1}j_2;J\rangle
\nonumber\\&&\qquad=
\sum_R
\langle(j_1j_2)J,(j_3j_4)J;0|(j_3j_2)R,(j_1j_4)R;0\rangle
\langle j_1j_4;R\|\hat V\|j_3j_2;R\rangle_{\rm a},
\label{e_bm1}
\end{eqnarray}
where the first symbol in angle brackets represents a re-coupling coefficient of four angular momenta
and the subscript `a' of the second symbol in angle brackets
indicates that the reduced matrix element
is taken between anti-symmetric two-particle states.
This relation implies
\begin{eqnarray}
&&\langle j_3^{-1}j_4;J|\hat V|j_1^{-1}j_2;J\rangle=
(-)^{j_1+j_2+j_3+j_4}
\langle j_2 j_1^{-1};J|\hat V|j_4j_3^{-1};J\rangle
\nonumber\\&&\qquad=
-\sum_R(2R+1)
\left\{\begin{array}{ccc}
j_1&j_2&J\\
j_3&j_4&R\end{array}\right\}
\langle j_1j_4;R|\hat V|j_3j_2;R\rangle_{\rm a},
\nonumber\\&&\qquad=
-\sum_R(2R+1)
\left\{\begin{array}{ccc}
j_1&j_2&J\\
j_3&j_4&R\end{array}\right\}
(-)^{j_1+j_2+j_3+j_4}
\langle j_4j_1;R|\hat V|j_2j_3;R\rangle_{\rm a},
\label{e_bm2}
\end{eqnarray}
and therefore
\begin{equation}
\langle j_1j_2^{-1};J|\hat V|j_3j_4^{-1};J\rangle=
-\sum_R(2R+1)
\left\{\begin{array}{ccc}
j_1&j_2&J\\
j_3&j_4&R\end{array}\right\}
\langle j_1j_4;R|\hat V|j_3j_2;R\rangle_{\rm a}.
\label{e_bm3}
\end{equation}
The requirement of anti-symmetry implies that,
if $j_1=j_4$ or $j_2=j_3$,
the summation in Eq.~(\ref{e_bm2}) is restricted to even values of $R$.
The result~(\ref{e_bm2}) agrees with the transformation used in Eq.~(\ref{e_vph1})
because the summation in that case is unrestricted
since particle-like (primed) and hole-like (unprimed) orbitals belong to different sets.
The result~(\ref{e_bm2}) also agrees with transformation used in Eq.~(\ref{e_vph2});
in that case the orbitals may be the same
but one orbital contains a neutron and the other a proton.

It is of interest to carry out an approximate diagonalisation
of the matrix with elements~(\ref{e_ham}) in two stages.
First, the like-nucleon interaction is diagonalised for neutrons and protons separately
with use of the expression~(\ref{e_vph1}),
leading to a neutron and a proton octupole phonon,
\begin{equation}
|3^-_{{\rm c}\nu}\rangle\equiv
\sum_{k'k}\tilde\nu_{k'k}|j_{\nu k'}j_{\nu k}^{-1};3^-\rangle,
\qquad
|3^-_{{\rm c}\pi}\rangle\equiv
\sum_{k'k}\tilde\pi_{k'k}|j_{\pi k'}j_{\pi k}^{-1};3^-\rangle,
\label{e_ophnp}
\end{equation}
where coefficients $\tilde\rho_{k'k}$ (with tilde) are used
to distinguish them from the coefficients $\rho_{k'k}$ in Eq.~(\ref{e_oph}),
which result from the diagonalisation of the Hamiltonian in the complete particle--hole space.
In the model proposed in Sect.~\ref{s_solvable},
the diagonalisation within the subspace spanned by the two octupole phonons~(\ref{e_ophnp})
is approximately decoupled from the rest of the particle--hole space.
Under this assumption the diagonalisation in the complete particle--hole space
reduces to the secular problem associated with the matrix
\begin{equation}
\left[\begin{array}{cc}
E(3^-_{{\rm c}\nu})&V_{\nu\pi}\\V_{\nu\pi}&E(3^-_{{\rm c}\pi})
\end{array}\right],
\label{e_mat2}
\end{equation}
where $E(3^-_{{\rm c}\rho})$ are the energies of neutron and proton octupole phonons,
and the off-diagonal matrix element $V_{\nu\pi}$
depends only on the neutron--proton interaction,
\begin{equation}
V_{\nu\pi}=
-\sum_{k'kl'l}\tilde\nu_{k'k}\tilde\pi_{l'l}
\sum_R(2R+1)
\left\{\begin{array}{ccc}
j_{\nu k'}&j_{\nu k}&3\\
j_{\pi l'}&j_{\pi l}&R\end{array}\right\}
V^{\nu\pi R}_{k'll'k}.
\label{e_offd}
\end{equation}

Let us dwell a little longer on the solution of Eq.~(\ref{e_mat2}).
Assume that the neutron octupole phonon is lowest in energy,
$E(3^-_{{\rm c}\nu})<E(3^-_{{\rm c}\pi})$;
the opposite case is obtained
by interchanging neutron and proton indices in the following.
The eigenenergies of the matrix~(\ref{e_mat2}) are given by
\begin{equation}
E(3^-_{{\rm c}\pm})=
E(3^-_{{\rm c}\nu})+\Delta E\pm\sqrt{\Delta E^2+V_{\nu\pi}^2},
\label{e_eige2}
\end{equation}
where $\Delta E=[E(3^-_{{\rm c}\pi})-E(3^-_{{\rm c}\nu})]/2$,
and the associated eigenfunctions can be written as
\begin{equation}
|3^-_{{\rm c}\pm}\rangle=
\frac{1}{\sqrt{2f_v(f_v\mp1)}}
\left[-(1\mp f_v)|3^-_{{\rm c}\nu}\rangle+v|3^-_{{\rm c}\pi}\rangle\right],
\label{e_eigw2}
\end{equation}
where $v=V_{\nu\pi}/\Delta E$ and $f_v=\sqrt{1+v^2}\geq1$.
One therefore predicts the occurrence of {\em two} collective $3^-$ states,
which, in the limit of large $|v|$, are the symmetric and the anti-symmetric combinations
of the neutron and proton octupole phonons, respectively, since
\begin{equation}
|3^-_{{\rm c}\pm}\rangle\stackrel{|v|\rightarrow\infty}{\longrightarrow}
\frac{1}{\sqrt{2v^2}}
\left[\pm|v||3^-_{{\rm c}\nu}\rangle+v|3^-_{{\rm c}\pi}\rangle\right]=
\frac{1}{\sqrt{2}}
\left[\pm|3^-_{{\rm c}\nu}\rangle+{\rm sign}(V_{\nu\pi})|3^-_{{\rm c}\pi}\rangle\right].
\label{e_eigw2b}
\end{equation}
The collective octupole phonon that is lowest in energy
can be symmetric or anti-symmetric in the neutron and proton octupole phonons,
depending on the sign of $V_{\nu\pi}$.
As will be shown in Sect.~\ref{s_solvable},
for a short-range, attractive neutron--proton interaction,
$V_{\nu\pi}$ is negative
and in that case the low-energy octupole excitation
is a symmetric combination of the neutron and proton octupole phonons.

Of particular interest is the strength of the transition
from the collective octupole phonon to the ground state,
which is indicated by the size of its $B(E3)$ value
\begin{equation}
B(E3;3^-_{\rm c}\rightarrow0^+_1)=
{\frac17}|\langle0^+_1\|\hat T(E3)\|3^-_{\rm c}\rangle|^2,
\label{e_be3}
\end{equation}
with
\begin{eqnarray}
\langle0^+_1\|\hat T(E3)\|3^-_{\rm c}\rangle&=&
\sum_{k'k}\nu_{k'k}\langle0^+_1\|e_\nu\hat T_\nu(E3)\|j_{\nu k'}j_{\nu k}^{-1};3^-\rangle
\nonumber\\&+&
\sum_{k'k}\pi_{k'k}\langle0^+_1\|e_\pi\hat T_\pi(E3)\|j_{\pi k'}j_{\pi k}^{-1};3^-\rangle,
\label{e_me3}
\end{eqnarray}
where $e_\rho$ is the effective charge for neutron or proton.
The evaluation of the $B(E3)$ value~(\ref{e_be3})
therefore requires the calculation of
 \begin{eqnarray}
\langle0^+_1\|\hat T(E\lambda)\|j'j^{-1};\lambda\rangle&=&
-(-)^{j'+j-\lambda}\langle0^+_1\|\hat T(E\lambda)\|j^{-1}j';\lambda\rangle
\nonumber\\&=&
(-)^{j'+j-\lambda}\langle j\|\hat T(E\lambda)\|j'\rangle,
\label{e_rmsp1}
\end{eqnarray}
where in the last line use is made of Eq.~(3B-25) of Ref.~\cite{BM69}.
The electric-transition operator of multipolarity $\lambda$
is $\hat T(E\lambda)=r^\lambda Y_{\lambda\mu}$,
in which case the reduced matrix element in Eq.~(\ref{e_rmsp1}) is obtained from
\begin{eqnarray}
&&\langle n\ell{\textstyle{\frac12}}j\|r^\lambda Y_\lambda\|n'\ell'{\textstyle{\frac12}}j'\rangle
\nonumber\\&&\qquad=
(-)^{j'-1/2}\sqrt{(2j+1)(2\lambda+1)(2j'+1)}
\left(\begin{array}{ccc}
j&\lambda&j'\\-{\frac12}&0&{\frac12}
\end{array}\right)b^\lambda I_{n\ell n'\ell'}^{(\lambda)},
\label{e_rmsp2}
\end{eqnarray}
where the symbol between round brackets is a Wigner $3j$ coefficient~\cite{Shalit63,Talmi93}
and where it is assumed that $(-)^{\lambda+\ell+\ell'}=+1$.
The length parameter $b$ characterises
the size of the harmonic-oscillator potential
and $I_{n\ell n'\ell'}^{(\lambda)}$ is the radial integral
\begin{equation}
I_{n\ell n'\ell'}^{(\lambda)}=
\frac{1}{\sqrt{4\pi}}\int_0^{+\infty}\left(\frac{r}{b}\right)^\lambda R_{n\ell}(r)R_{n'\ell'}(r)r^2dr.
\label{e_radint1}
\end{equation}
In the phase convention of radial wave functions $R_{n\ell}(r)$ that are positive for $r\rightarrow\infty$,
the radial integral~(\ref{e_radint1}) is the {\em positive} quantity
\begin{eqnarray}
I_{n\ell n'\ell'}^{(\lambda)}&=&
\sqrt{\frac{n!n'!}{4\pi\Gamma(n+\ell+3/2)\Gamma(n'+\ell'+3/2)}}\Gamma(p+1)
\nonumber\\&&\times
\sum_k
\left(\begin{array}{c}p-\ell-1/2\\n-k\end{array}\right)
\left(\begin{array}{c}p-\ell'-1/2\\n'-k\end{array}\right)
\left(\begin{array}{c}p+k\\k\end{array}\right),
\label{e_radint2}
\end{eqnarray}
with $p=(\ell+\ell'+\lambda+1)/2$.

\section{A solvable model for octupole phonons}
\label{s_solvable}
\begin{figure}
\resizebox{0.5\columnwidth}{!}{\includegraphics{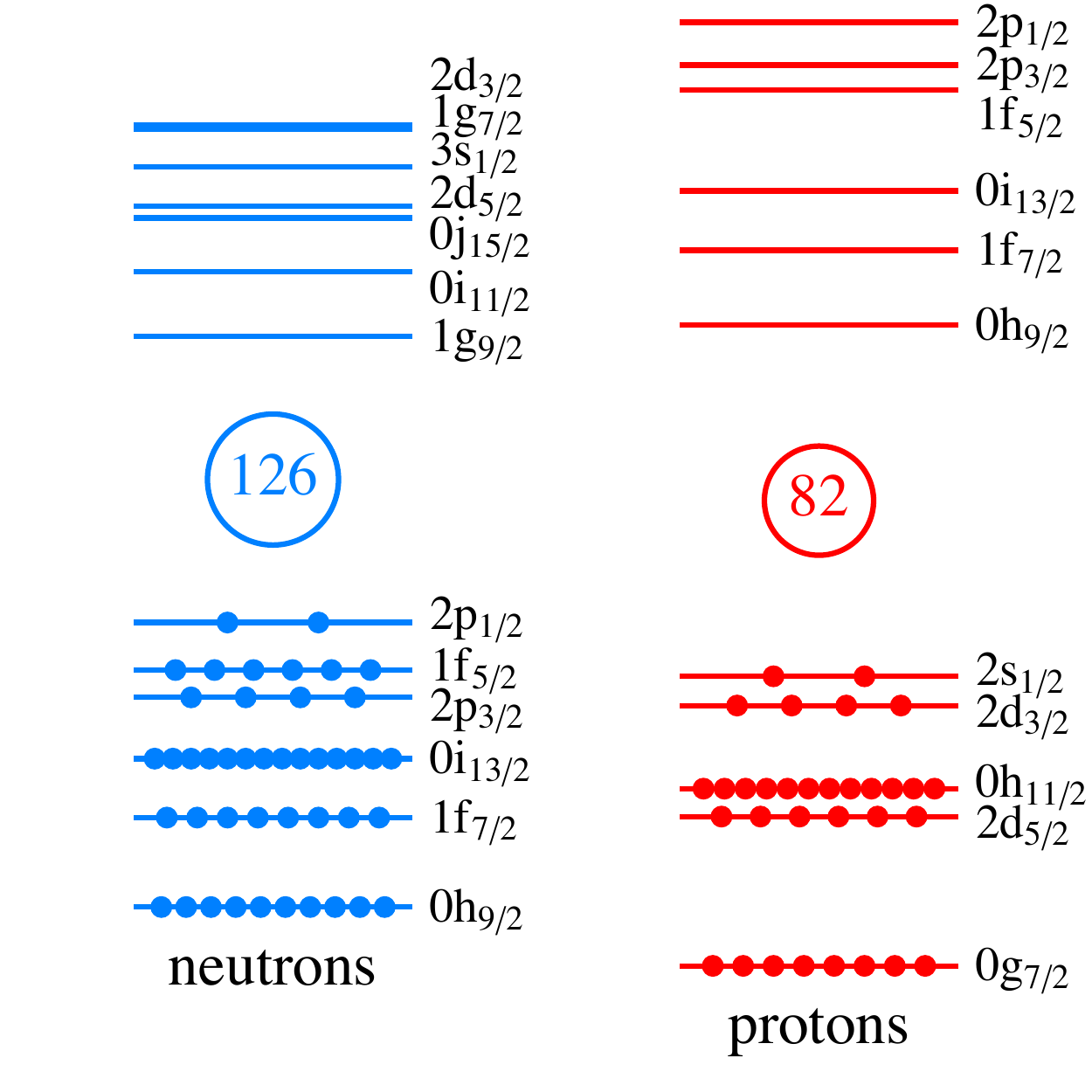} }
\resizebox{0.5\columnwidth}{!}{\includegraphics{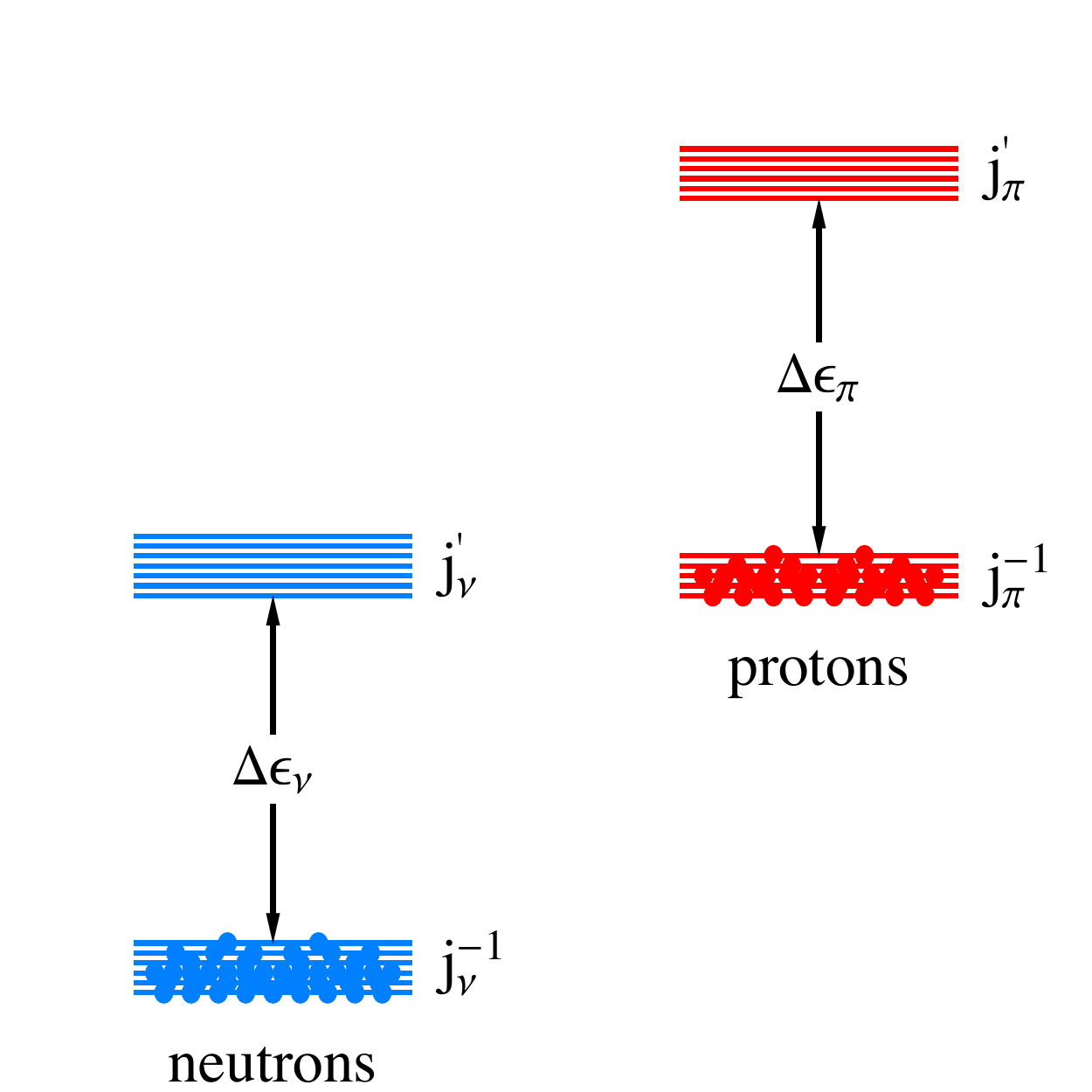} }
\caption{Single-particle energies
as taken in a realistic shell-model calculation for $^{208}$Pb (left)
and as assumed in the solvable model (right).}
\label{f_spe}
\end{figure}
Let us study the features of octupole phonons in a simplified model,
which assumes degenerate single-particle energies below and above the shell closures,
such that
\begin{equation}
\epsilon_{\rho k'}-\epsilon_{\rho k}=\Delta\epsilon_\rho,
\qquad\forall k,k',
\label{e_degen}
\end{equation}
as illustrated on the right-hand side of Fig.~\ref{f_spe}.
Furthermore, it is assumed that a surface delta interaction (SDI) acts between the nucleons,
whose matrix elements in the isospin formalism are~\cite{Brussaard77}
\begin{eqnarray}
&&\langle j_1j_2;JT|\hat V_{\rm SDI}|j_3j_4;JT\rangle\equiv
\langle n_1\ell_1j_1,n_2\ell_2j_2;JT|\hat V_{\rm SDI}|n_3\ell_3j_3,n_4\ell_4j_4;JT\rangle
\label{e_sdi}\\&&\qquad=
a_TF
\left\{
F'\left(\begin{array}{ccc}
j_1&j_2&J\\\frac12&-\frac12&0
\end{array}\right)
\left(\begin{array}{ccc}
j_3&j_4&J\\\frac12&-\frac12&0
\end{array}\right)-
\delta_{T0}
\left(\begin{array}{ccc}
j_1&j_2&J\\\frac12&\frac12&-1
\end{array}\right)
\left(\begin{array}{ccc}
j_3&j_4&J\\\frac12&\frac12&-1
\end{array}\right)
\right\},
\nonumber
\end{eqnarray}
in terms of the (positive) isoscalar and isovector strength parameters, $a_0$ and $a_1$,
and where the following factors are introduced:
\begin{eqnarray}
F&\equiv&\sqrt{\frac{(2j_1+1)(2j_2+1)(2j_3+1)(2j_4+1)}{(1+\delta_{12})(1+\delta_{34})}},
\nonumber\\
F'&\equiv&(-)^{j_2+j_4+\ell_2+\ell_4}{\frac12}\left[1-(-)^{J+T+\ell_3+\ell_4}\right].
\label{e_factors}
\end{eqnarray}
Note the difference of an overall phase $(-)^{n_1+n_2+n_3+n_4}$
with respect to Eq.~(6.43) of Ref.~\cite{Brussaard77},
due to the different phase convention for the radial wave functions.

Consider first the neutron and proton octupole phonons~(\ref{e_ophnp}) separately.
Their structure depends solely on the $T=1$ matrix elements,
which for the SDI reduce to
\begin{equation}
\langle j_1j_2;JT=1|\hat V_{\rm SDI}|j_3j_4;JT=1\rangle=
a_1FF'
\left(\begin{array}{ccc}
j_1&j_2&J\\\frac12&-\frac12&0
\end{array}\right)
\left(\begin{array}{ccc}
j_3&j_4&J\\\frac12&-\frac12&0
\end{array}\right).
\label{e_sdi1}
\end{equation}
This matrix element must be introduced
in the expression~(\ref{e_vph1}) for the particle--hole matrix element,
which requires the evaluation of sums summarised in Appendix~\ref{a_sums1}.
As a result, for a state with natural parity, that is, with $(-)^{\ell_{\rho k}+\ell_{\rho k'}+J}=+1$,
the particle--hole matrix element~(\ref{e_vph1}) can be written as
\begin{eqnarray}
\langle j_{\rho k'}j_{\rho k}^{-1};J|\hat H_\rho|j_{\rho l'}j_{\rho l}^{-1};J\rangle&=&
\Delta\epsilon_\rho\delta_{k'k}\delta_{l'l}+
\langle j_{\rho k'}j_{\rho k}^{-1};J|\hat V_{\rho\rho}^{\rm SDI}|j_{\rho l'}j_{\rho l}^{-1};J\rangle
\nonumber\\&=&
\Delta\epsilon_\rho\delta_{k'k}\delta_{l'l}+
\frac{a_{1\rho}}{2}
\bigl(f^\rho_{k'k}f^\rho_{l'l}-g^\rho_{k'k}g^\rho_{l'l}\bigr),
\label{e_sdiph1}
\end{eqnarray}
where
\begin{eqnarray}
f^\rho_{k'k}&=&
(-)^{\ell_{\rho k}}\sqrt{(2j_{\rho k}+1)(2j_{\rho k'}+1)}
\left(\begin{array}{ccc}
j_{\rho k'}&j_{\rho k}&J\\\frac12&\frac12&-1
\end{array}\right),
\nonumber\\
g^\rho_{k'k}&=&
(-)^{j_{\rho k}-1/2}\sqrt{(2j_{\rho k}+1)(2j_{\rho k'}+1)}
\left(\begin{array}{ccc}
j_{\rho k'}&j_{\rho k}&J\\\frac12&-\frac12&0
\end{array}\right).
\label{e_deffg}
\end{eqnarray}
The Hamiltonian matrix~(\ref{e_sdiph1})
can thus be written as the sum of two separable matrices,
leading to a two-dimensional subspace
that decouples {\em exactly} from the complete particle--hole space
and to a secular equation in terms of a $2\times2$ matrix of the form
\begin{equation}
\frac{a_{1\rho}}{2}
\left[\begin{array}{cc}
\displaystyle\sum_{k'k}(f^\rho_{k'k})^2&\imath\displaystyle\sum_{k'k}f^\rho_{k'k}g^\rho_{k'k}\\
\imath\displaystyle\sum_{k'k}f^\rho_{k'k}g^\rho_{k'k}&\displaystyle-\sum_{k'k}(g^\rho_{k'k})^2
\end{array}\right].
\label{e_sdih2}
\end{equation}
The eigenvalues of this matrix correspond {\em exactly}
to the two non-zero eigenvalues in the complete space.
Furthermore, in most applications the off-diagonal matrix element in Eq.~(\ref{e_sdih2})
is small compared with the diagonal matrix elements
because of a property of Wigner $3j$ coefficients (see Appendix~\ref{a_sums2}),
\begin{equation}
\sum_{k'k}f^\rho_{k'k}g^\rho_{k'k}\ll
\sum_{k'k}(f^\rho_{k'k})^2+\sum_{k'k}(g^\rho_{k'k})^2.
\label{e_ggfg1}
\end{equation}
The following approximation can therefore be made
for the energy and wave function of the neutron and proton octupole phonons:
\begin{equation}
E(3^-_{{\rm c}\rho})\approx
\Delta\epsilon_\rho-\frac{a_{1\rho}}{2}S_\rho=
\Delta\epsilon_\rho-\frac{a_{1\rho}}{2}\sum_{k'k}(2j_{\rho k}+1)(2j_{\rho k'}+1)
\left(\begin{array}{ccc}
j_{\rho k'}&j_{\rho k}&3\\\frac12&-\frac12&0
\end{array}\right)^2,
\label{e_sdien}
\end{equation}
and
\begin{equation}
\tilde\rho_{k'k}\approx
\frac{g^\rho_{k'k}}{\sqrt{S_\rho}}=
(-)^{j_{\rho k}-1/2}
\sqrt{\frac{(2j_{\rho k}+1)(2j_{\rho k'}+1)}{S_\rho}}
\left(\begin{array}{ccc}
j_{\rho k'}&j_{\rho k}&3\\\frac12&-\frac12&0
\end{array}\right),
\label{e_sdiwf}
\end{equation}
where $S_\rho$ is the sum
\begin{equation}
S_\rho=
\sum_{k'k}\left(g^\rho_{k'k}\right)^2=
\sum_{k'k}(2j_{\rho k}+1)(2j_{\rho k'}+1)
\left(\begin{array}{ccc}
j_{\rho k'}&j_{\rho k}&3\\\frac12&-\frac12&0
\end{array}\right)^2.
\label{e_norm}
\end{equation}

Equation~(\ref{e_sdien}) gives an approximate expression for the diagonal matrix elements
in the $2\times2$ matrix~(\ref{e_mat2}).
The remaining problem is to calculate the off-diagonal matrix element $V_{\nu\pi}$,
originating from the neutron--proton interaction,
which for the SDI has the following matrix elements
\begin{eqnarray}
&&\langle j_{\nu k'}j_{\pi l};J|\hat V_{\nu\pi}^{\rm SDI}|j_{\nu k}j_{\pi l'};J\rangle
\label{e_sdinp}\\&&\qquad=
{\frac12}F
\left\{
b_1
\left(\begin{array}{ccc}
j_{\nu k'}&j_{\pi l}&J\\\frac12&-\frac12&0
\end{array}\right)
\left(\begin{array}{ccc}
j_{\nu k}&j_{\pi l'}&J\\\frac12&-\frac12&0
\end{array}\right)+
b_0
\left(\begin{array}{ccc}
j_{\nu k'}&j_{\pi l}&J\\\frac12&\frac12&-1
\end{array}\right)
\left(\begin{array}{ccc}
j_{\nu k}&j_{\pi l'}&J\\\frac12&\frac12&-1
\end{array}\right)
\right\},
\nonumber
\end{eqnarray}
with
\begin{equation}
b_1=
(-)^{j_{\pi l}+j_{\pi l'}+\ell_{\pi l}+\ell_{\pi l'}}
\left[\frac{a_1+a_0}{2}+(-)^{J+\ell_{\nu k}+\ell_{\pi l'}}\frac{a_1-a_0}{2}\right],
\quad
b_0=-a_0.
\label{e_bcoef}
\end{equation}
If this expression is introduced in the particle--hole matrix element~(\ref{e_vph2}),
the resulting sums can be evaluated (see Appendix~\ref{a_sums1}), leading to
\begin{eqnarray}
&&\langle j_{\nu k'}j_{\nu k}^{-1};J|\hat V_{\nu\pi}^{\rm SDI}|j_{\pi l'}j_{\pi l}^{-1};J\rangle
\label{e_sdinpph}\\&&\qquad=
{\frac12}F
\left\{
b'_1
\left(\begin{array}{ccc}
j_{\nu k'}&j_{\nu k}&J\\\frac12&-\frac12&0
\end{array}\right)
\left(\begin{array}{ccc}
j_{\pi l'}&j_{\pi l}&J\\\frac12&-\frac12&0
\end{array}\right)+
b'_0
\left(\begin{array}{ccc}
j_{\nu k'}&j_{\nu k}&J\\\frac12&\frac12&-1
\end{array}\right)
\left(\begin{array}{ccc}
j_{\pi l'}&j_{\pi l}&J\\\frac12&\frac12&-1
\end{array}\right)
\right\},
\nonumber
\end{eqnarray}
with
\begin{equation}
b'_1=
(-)^{j_{\nu k}+j_{\pi l}}
\left[(-)^{J+\ell_{\pi l}+\ell_{\pi l'}}\frac{a_1+a_0}{2}+a_0\right],
\quad
b'_0=(-)^{\ell_{\nu k}+\ell_{\pi l}}\frac{a_1-a_0}{2}.
\label{e_bpcoef}
\end{equation}
For a state with natural parity, $(-)^{\ell_{\nu k}+\ell_{\nu k'}+J}=(-)^{\ell_{\pi l}+\ell_{\pi l'}+J}=+1$,
the matrix element reduces to
\begin{equation}
\langle j_{\nu k'}j_{\nu k}^{-1};J|\hat V_{\nu\pi}^{\rm SDI}|j_{\pi l'}j_{\pi l}^{-1};J\rangle=
\frac{a_1-a_0}{4}f^\nu_{k'k}f^\pi_{l'l}-
\frac{a_1+3a_0}{4}g^\nu_{k'k}g^\pi_{l'l}.
\label{e_sdiph2}
\end{equation}
In a final step the particle--hole matrix element~(\ref{e_sdiph2})
is introduced in the expansion~(\ref{e_offd})
with use of the coefficients~(\ref{e_sdiwf}),
which are approximately valid for a SDI.
The sums to be evaluated have the property (see Appendix~\ref{a_sums2})
\begin{equation}
\sum_{k'kl'l}f^\nu_{k'k}g^\nu_{k'k}f^\pi_{l'l}g^\pi_{l'l}\ll
\sum_{k'kl'l}g^\nu_{k'k}g^\nu_{k'k}g^\pi_{l'l}g^\pi_{l'l},
\label{e_ggfg2}
\end{equation}
such that, to a good approximation, one obtains
\begin{equation}
V_{\nu\pi}\approx-\frac{a_1+3a_0}{4}\sqrt{S_\nu S_\pi}.
\label{e_offda}
\end{equation}

In summary,
if one assumes degenerate single-particle energies below and above the shell closures
and if a SDI among the nucleons is taken,
the energies of the symmetric and anti-symmetric octupole phonons are,
to a good approximation,
obtained from the $2\times2$ matrix
\begin{equation}
\left[\begin{array}{cc}
\displaystyle
\Delta\epsilon_\nu-\frac{a_{1\nu}}{2}S_\nu&
\displaystyle
-\frac{a_1+3a_0}{4}\sqrt{S_\nu S_\pi}\\
\displaystyle
-\frac{a_1+3a_0}{4}\sqrt{S_\nu S_\pi}&
\displaystyle
\Delta\epsilon_\pi-\frac{a_{1\pi}}{2}S_\pi
\end{array}\right].
\label{e_mat2a}
\end{equation}
Since the isoscalar and isovector strengths $a_0$ and $a_1$ are positive,
the off-diagonal element of this matrix is negative
and therefore the low-energy collective $3^-$ state
is the (approximately) symmetric combination of the neutron and proton octupole phonons,
which in general can be written as
\begin{equation}
|3^-_{\rm c}\rangle=
\alpha_\nu|3^-_{{\rm c}\nu}\rangle+\alpha_\pi|3^-_{{\rm c}\pi}\rangle,
\label{e_eigw2c}
\end{equation}
where $\alpha_\nu$ and $\alpha_\pi$ have the same sign.

For the $E3$ transition strength one finds under the same assumptions
\begin{eqnarray}
\langle0^+_1\|\hat T_\rho(E3)\|3^-_{{\rm c}\rho}\rangle&=&
\sum_{k'k}\rho_{k'k}\langle0^+_1\|\hat T_\rho(E3)\|j_{\rho k'}j_{\rho k}^{-1};3^-\rangle
\nonumber\\&\approx&
\sum_{k'k}\tilde\rho_{k'k}\langle0^+_1\|\hat T_\rho(E3)\|j_{\rho k'}j_{\rho k}^{-1};3^-\rangle=
\frac{S^{(3)}_\rho}{\sqrt{S_\rho}}b^3,
\label{e_me3a}
\end{eqnarray}
where
\begin{equation}
S^{(\lambda)}_\rho\equiv
\sqrt{2\lambda+1}
\sum_{k'k}(2j_{\rho k}+1)(2j_{\rho k'}+1)
\left(\begin{array}{ccc}
j_{\rho k'}&j_{\rho k}&\lambda\\\frac12&-\frac12&0
\end{array}\right)^2
I_{n_{\rho k}\ell_{\rho k}n_{\rho k'}\ell_{\rho k'}}^{(\lambda)}.
\label{e_norml}
\end{equation}
The total $E3$ transition strength
from the low-energy collective octupole excitation $3^-_{\rm c}$ to the ground state
is therefore
\begin{equation}
\langle0^+_1\|\hat T(E3)\|3^-_{\rm c}\rangle=
\left(e_\nu\alpha_\nu\frac{S^{(3)}_\nu}{\sqrt{S_\nu}}+
e_\pi\alpha_\pi\frac{S^{(3)}_\pi}{\sqrt{S_\pi}}\right)b^3.
\label{e_me3at}
\end{equation}
All contributions in this expression add coherently.
This is so for the terms appearing in the sum~(\ref{e_norml})
because, as remarked earlier, the phase convention is such
that all radial integrals $I_{n\ell n'\ell'}^{(\lambda)}$  are positive.
Furthermore, the neutron and proton contributions in Eq.~(\ref{e_me3at}) add coherently
because the $3^-_{\rm c}$ state is the symmetric combination
of the neutron and proton octupole phonons,
implying that $\alpha_\nu$ and $\alpha_\pi$ have the same sign.

\section{An application to $^{208}$Pb}
\label{s_pb208}
Let us now investigate to what extent the simple properties
of the collective octupole state in the solvable model of the previous section
are found in a shell-model calculation
with a realistic single-particle space and two-body interaction.
As an example we consider the nucleus $^{208}$Pb.

The single-particle orbitals in the application presented in this section
span two major oscillator shells.
For the neutrons they include the orbitals below the $N=126$ shell closure,
$2p_{1/2}$, $2p_{3/2}$, $1f_{5/2}$, $1f_{7/2}$, $0h_{9/2}$ and $0i_{13/2}$,
and the ones above,
$3s_{1/2}$, $2d_{3/2}$, $2d_{5/2}$, $1g_{7/2}$, $1g_{9/2}$, $0i_{11/2}$ and $0j_{15/2}$;
for the protons they include the orbitals below the $Z=82$ shell closure,
$2s_{1/2}$, $1d_{3/2}$, $1d_{5/2}$, $0g_{7/2}$ and $0h_{11/2}$,
and the ones above, 
$2p_{1/2}$, $2p_{3/2}$, $1f_{5/2}$, $1f_{7/2}$, $0h_{9/2}$ and $0i_{13/2}$.

\begin{figure}
\centering
\resizebox{0.8\columnwidth}{!}{\includegraphics{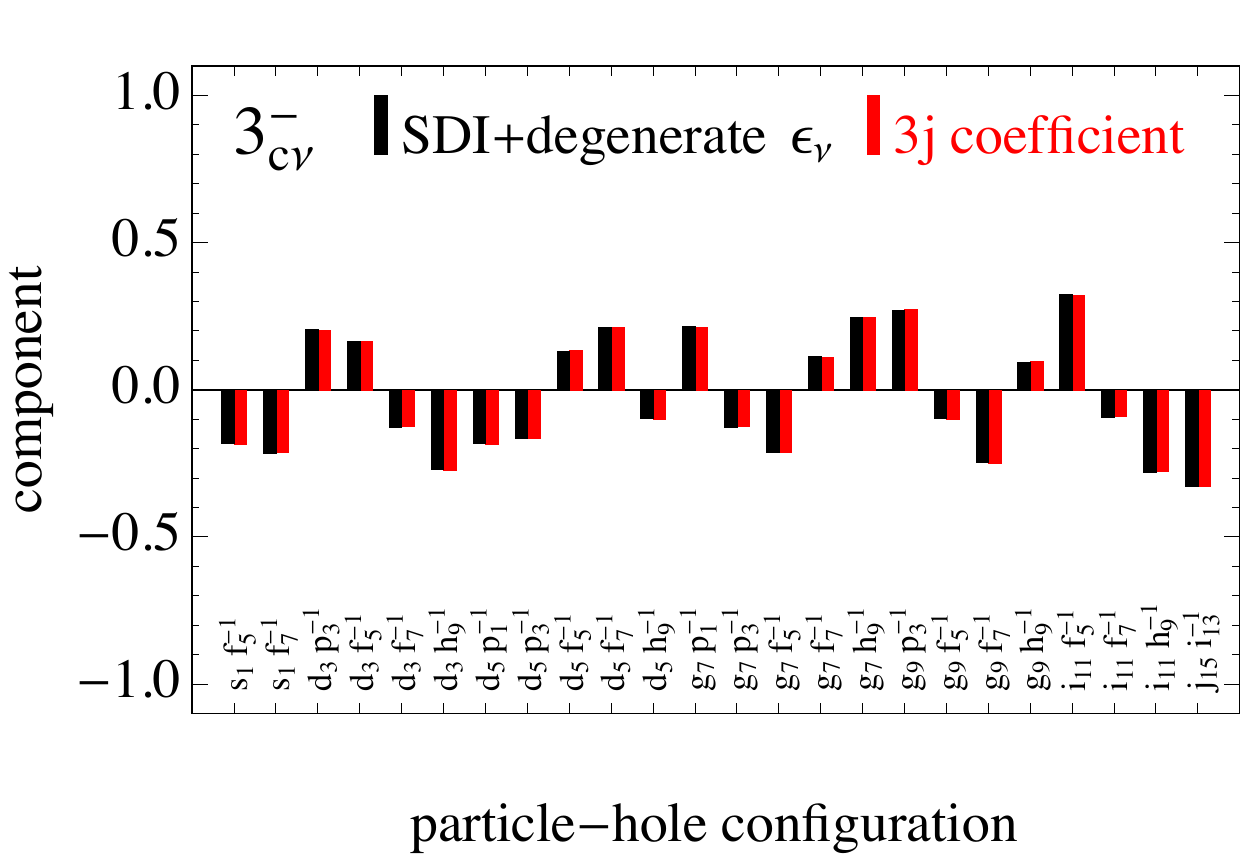} }
\resizebox{0.8\columnwidth}{!}{\includegraphics{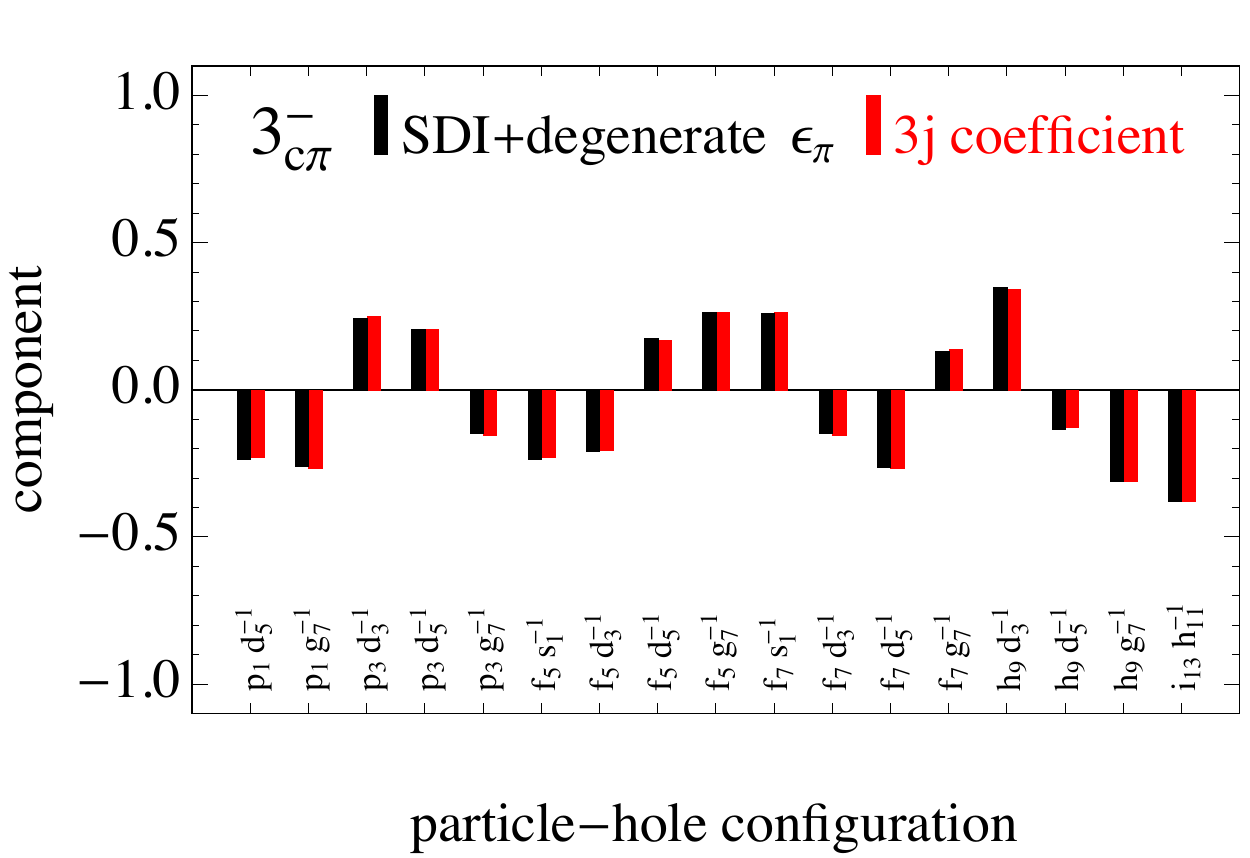} }
\caption{The particle--hole components of the wave function
of the low-energy collective $3^-_{{\rm c}\rho}$ states for neutrons (top) and protons (bottom), 
calculated numerically (black) and analytically from Eq.~(\ref{e_sdiwf}) (red).
The single-particle space is appropriate for $^{208}$Pb (see text).
The numerical results are obtained
with degenerate single-particle energies and with a SDI among the nucleons.
On the $x$-axis are indicated all possible configurations in a simplified notation,
e.g.\ $s_1f_5^{-1}$ stands for $3s_{1/2}1f_{5/2}^{-1}$.}
\label{f_wfsdi}
\end{figure}
For this heavy nucleus the major shells
obviously do not coincide with those of a harmonic oscillator
and display intruding and extruding orbitals.
As a result, the off-diagonal element in the matrix~(\ref{e_sdih2})
is not exactly zero (see Appendix~\ref{a_sums2})
and a first question that arises is to what extent
the analytic components~(\ref{e_sdiwf})
in terms of Wigner $3j$ coefficients are valid.
This is illustrated in Fig.~\ref{f_wfsdi},
which compares the expression~(\ref{e_sdiwf})
with the results of a numerical calculation
using the above single-particle space
with degenerate single-particle energies and a SDI among the nucleons.
While the energy of $3^-_{{\rm c}\rho}$ state depends
on the splitting $\Delta\epsilon_\rho$ and the strength $a_{1\rho}$ of the SDI,
its wave function is parameter free and determined by the matrix~(\ref{e_sdih2}).
The numerically calculated wave function
therefore only depends on the choice of single-particle space.
Furthermore, the correspondence between the wave function obtained numerically
and the analytic components in terms of Wigner $3j$ coefficients
is seen to be excellent,
which proves that the approximation~(\ref{e_ggfg1})
applies to a realistic choice of the single-particle space for $^{208}$Pb.

\begin{figure}
\centering
\resizebox{0.8\columnwidth}{!}{\includegraphics{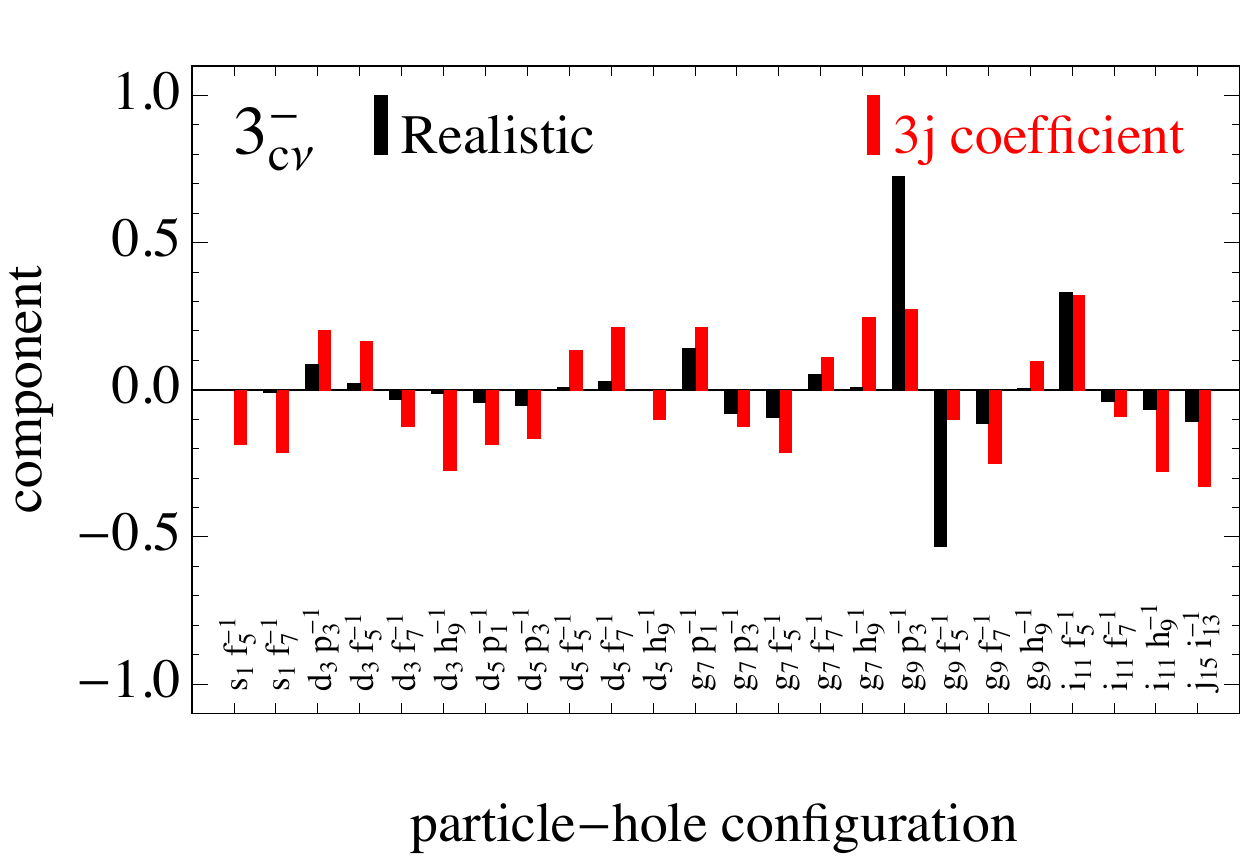} }
\resizebox{0.8\columnwidth}{!}{\includegraphics{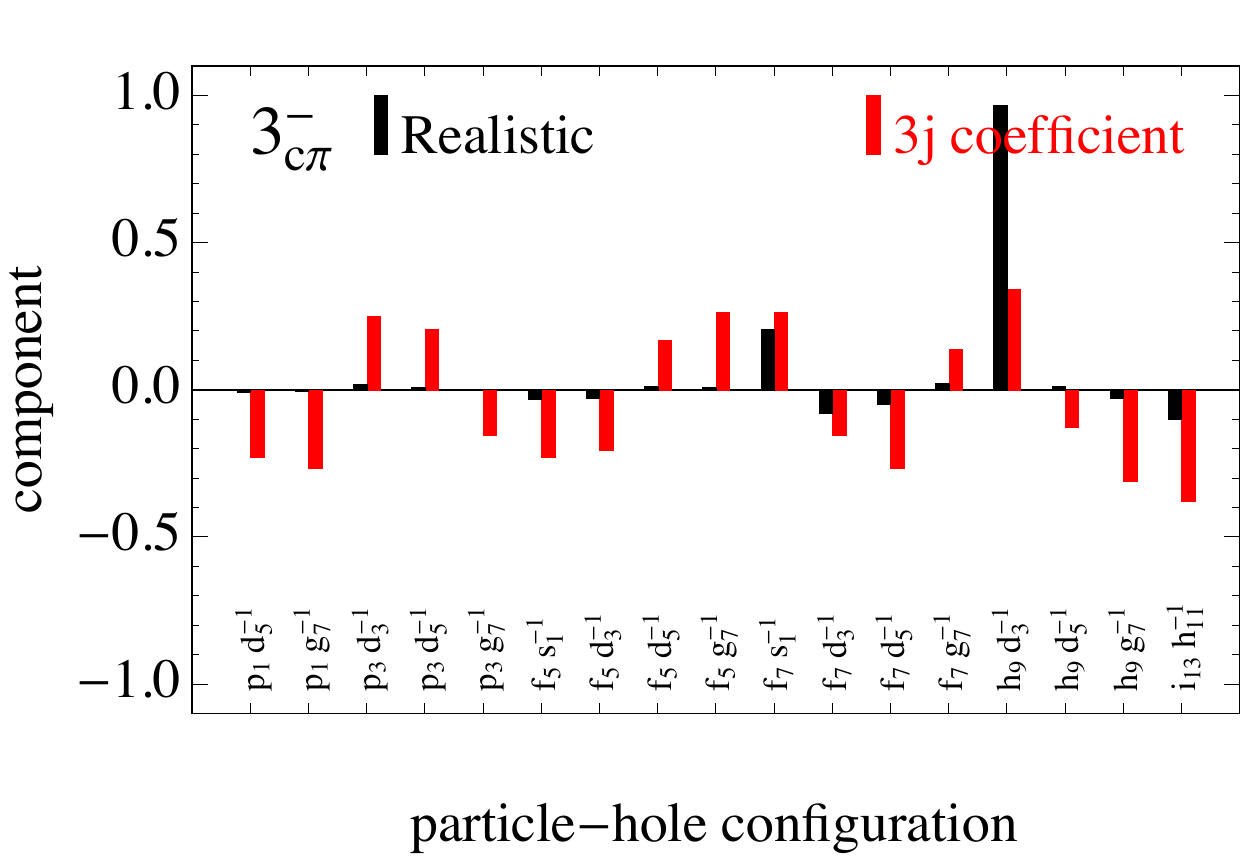} }
\caption{The particle--hole components of the wave function
of the low-energy collective $3^-_{{\rm c}\rho}$ states for neutrons (top) and protons (bottom), 
calculated numerically (black) and analytically from Eq.~(\ref{e_sdiwf}) (red).
The numerical results are obtained
with realistic single-particle energies
and two-body interaction appropriate for $^{208}$Pb (see text).
On the $x$-axis are indicated all possible configurations in a simplified notation,
e.g.\ $s_1f_5^{-1}$ stands for $3s_{1/2}1f_{5/2}^{-1}$.}
\label{f_wfreal}
\end{figure}
Nevertheless, actual single-particle energies are not degenerate
nor do the SDI two-body matrix elements coincide with those taken in a realistic calculation.
The single-particle energies appropriate for the $^{208}$Pb region
have been deduced from the data by Rejmund {\it et al.}~\cite{Rejmund99}
and these are shown on the right-hand side of Fig.~\ref{f_spe}.
There are about 35000 two-body matrix elements in this single-particle space,
which can be obtained in a variety of ways, as described by Brown~\cite{Brown00}.
The set used in the present calculation is taken from Refs.~\cite{Wrzesinski01,Rejmund20un}.
Figure~\ref{f_wfreal} compares the wave function components
of the neutron and proton octupole states obtained in a realistic calculation
with the analytic expression in terms of Wigner $3j$ coefficients.
The main conclusion from these results is
that much of the collectivity vanishes from the lowest $3^-$ excitations
since, in particular for the protons, the wave function
is dominated by only a few components.

\begin{figure}
\centering
\resizebox{0.8\columnwidth}{!}{\includegraphics{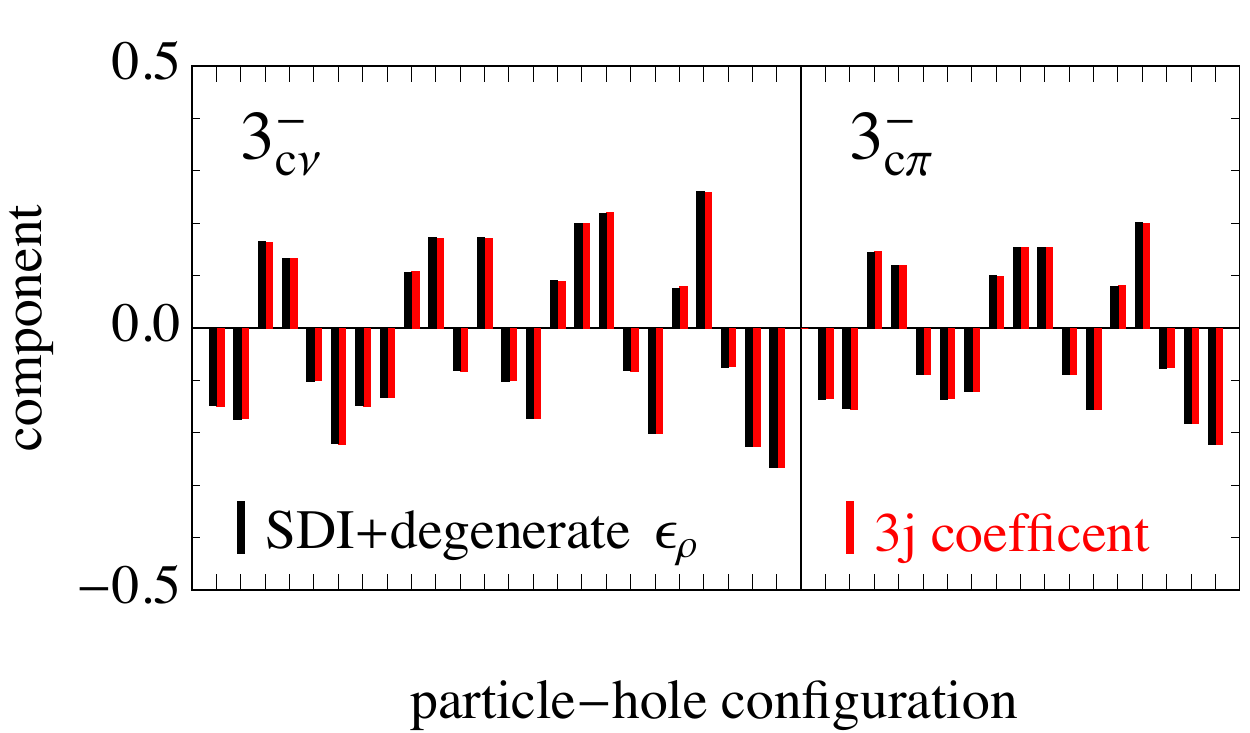} }
\resizebox{0.8\columnwidth}{!}{\includegraphics{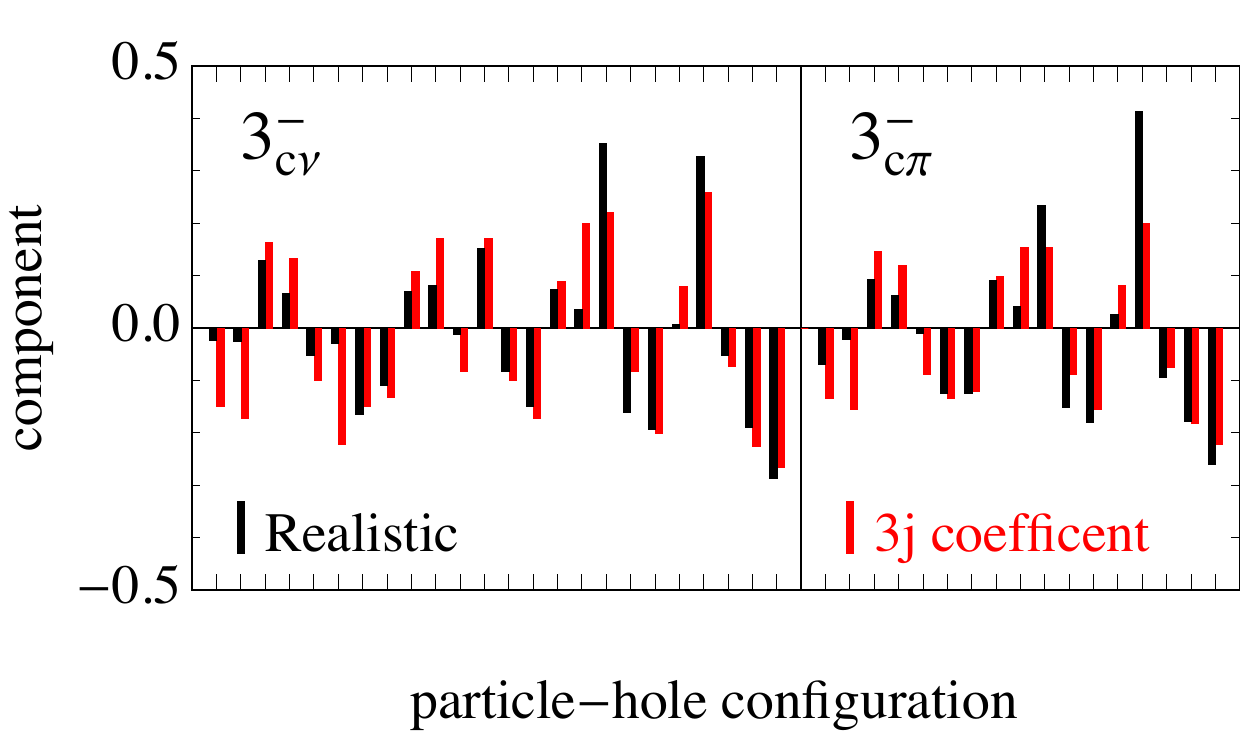} }
\caption{The particle--hole components of the wave function
of the low-energy collective $3^-_{\rm c}$ state,
calculated numerically (black) and analytically from Eqs.~(\ref{e_sdiwf}) and~(\ref{e_eigw2c}) (red).
The top panel compares the analytic expression
with a numerical calculation with degenerate single-particle energies
and a SDI among the nucleons.
The bottom panel compares the same analytic expression
with the results of a realistic shell-model calculation
as described in the text.}
\label{f_wf}
\end{figure}
It must be emphasised, however,
that the components of the neutron and proton octupole states as shown in Fig.~\ref{f_wfreal}
result from the neutron--neutron and proton--proton interactions only.
They correspond to the coefficients $\tilde\rho_{k'k}$
as obtained in the two-stage diagonalisation,
which, as explained in Sect.~\ref{s_general},
may be different from the coefficients $\rho_{k'k}$,
resulting from the diagonalisation
of the Hamiltonian in the complete particle--hole space.
The latter are shown in Fig.~\ref{f_wf}.
The top panel compares the components,
obtained from Eq.~(\ref{e_eigw2c}) together with the analytic expression~(\ref{e_sdiwf}),
with those from a numerical calculation with degenerate single-particle energies
and a SDI among the nucleons.
It displays therefore the same information as in Fig.~\ref{f_wfsdi}
but with each $3^-_{{\rm c}\rho}$ component multiplied with $\alpha_\rho$
as obtained from the diagonalisation of the $2\times2$ matrix~(\ref{e_mat2a}).
The bottom panel of Fig.~\ref{f_wf} compares the same analytic expression
with the components resulting from the realistic shell-model calculation.
It is clear that some of the components are enhanced while others are suppressed
as compared to the simple prescription in terms of a Wigner $3j$ coefficient.
Nevertheless, it should be noted that the phases of the realistic particle--hole wave function,
which constitute an important aspect of the collective octupole structure,
are in complete agreement with those predicted by the Wigner $3j$ coefficient.

\section{Summary and outlook}
\label{s_summ}
The ingredients of the solvable model for octupole phonons
in doubly-closed-shell nuclei are
(i) degenerate single-particle energies above and below the shell closures
and (ii) a surface delta interaction (SDI) among the nucleons.
With these assumptions it follows that the two-dimensional space
spanned by the collective neutron and proton octupole excitations
to a very good approximation decouples from the full particle--hole space.
The resulting neutron and proton octupole phonons are strongly coupled
by the neutron--proton interaction,
producing a symmetric combination at low energy
and an anti-symmetric one (sometimes referred to as isovector) at higher energy.
The appealing property of this simplified analysis
is that the wave-function components of the neutron and proton octupole phonons
are essentially given by Wigner $3j$ coefficients.

Confronted with the results of a shell-model calculation
with realistic single-particle energies and two-body interaction matrix elements,
differences with the above solvable model are revealed.
In particular, the neutron--neutron and proton--proton interactions by themselves
do not lead to collective neutron and proton octupole excitations
since much of the collectivity disappears from the low-energy $3^-$ states
mainly due to the non-degeneracy of the single-particle levels.
Therefore, the decoupling property mentioned above
is not confirmed by a realistic shell-model calculation.
However, when all interactions are active,
including the neutron--proton interaction,
octupole collectivity is regained
and the simple structure of the solvable model is approximately recovered.
One can therefore paradoxically claim
that the separate neutron and proton octupole phonons
exist by virtue of the neutron--proton interaction,
which, besides generating their collective structure,
also couples them.

The proposed solvable octupole model
might be of use in the study of more complex systems
or as a guidance to answer more intricate questions.
An example is the analysis of odd-mass nuclei
as regards the coupling of a particle or hole to an octupole phonon,
as was recently discussed for $^{207}$Pb~\cite{Ralet19}.
The issue of double octupole phonons
may provide another example
since the solvable octupole model might be amenable
to an extension to two-particle--two-hole excitations.
These problems are currently under study.

\section*{Acknowledgements}
\label{s_thanks}
I wish to thank Emmanuel Cl\'ement, Antoine Lemasson and Maurycy Rejmund
for raising my interest in this problem
and for many fruitful discussions,
and Maurycy Rejmund for providing the realistic shell-model
single-particle energies and two-body matrix elements.

\appendix
\section{Sums of $3j$ and $6j$ coefficients}
\label{a_sums1}
In the treatment of the Pandya transformation of the SDI
the following sums are needed:
\begin{eqnarray}
S_0&\equiv&
\sum_R(2R+1)
\left(\begin{array}{ccc}
j_{k'}&j_l&R\\\frac12&-\frac12&0
\end{array}\right)
\left(\begin{array}{ccc}
j_{l'}&j_k&R\\\frac12&-\frac12&0
\end{array}\right)
\left\{\begin{array}{ccc}
j_{k'}&j_k&J\\
j_{l'}&j_l&R\end{array}\right\},
\nonumber\\
\tilde S_0&\equiv&
\sum_R(-)^R(2R+1)
\left(\begin{array}{ccc}
j_{k'}&j_l&R\\\frac12&-\frac12&0
\end{array}\right)
\left(\begin{array}{ccc}
j_{l'}&j_k&R\\\frac12&-\frac12&0
\end{array}\right)
\left\{\begin{array}{ccc}
j_{k'}&j_k&J\\
j_{l'}&j_l&R\end{array}\right\},
\nonumber\\
S_1&\equiv&
\sum_R(2R+1)
\left(\begin{array}{ccc}
j_{k'}&j_l&R\\\frac12&\frac12&-1
\end{array}\right)
\left(\begin{array}{ccc}
j_{l'}&j_k&R\\\frac12&\frac12&-1
\end{array}\right)
\left\{\begin{array}{ccc}
j_{k'}&j_k&J\\
j_{l'}&j_l&R\end{array}\right\}.
\label{e_sums1}
\end{eqnarray}
With use of Eq.~(15.14) of Ref.~\cite{Shalit63}
these sums reduce to
\begin{eqnarray}
S_0=S_1&=&
-(-)^{j_k+j_l}\left(\begin{array}{ccc}
j_{k'}&j_k&J\\\frac12&\frac12&-1
\end{array}\right)
\left(\begin{array}{ccc}
j_{l'}&j_l&J\\\frac12&\frac12&-1
\end{array}\right),
\nonumber\\
\tilde S_0&=&
-(-)^J
\left(\begin{array}{ccc}
j_{k'}&j_k&J\\\frac12&-\frac12&0
\end{array}\right)
\left(\begin{array}{ccc}
j_{l'}&j_l&J\\\frac12&-\frac12&0
\end{array}\right).
\label{e_sums2}
\end{eqnarray}

\section{Properties of the $2\times2$ matrix~(\ref{e_sdih2})}
\label{a_sums2}
The $2\times2$ matrix~(\ref{e_sdih2}) in the model
with degenerate single-particle energies and a SDI among the nucleons
has the following elements:
\begin{equation}
\sum_{k'k}f^\rho_{k'k}g^\rho_{k'k},
\qquad
\sum_{k'k}(f^\rho_{k'k})^2,
\qquad
\sum_{k'k}(g^\rho_{k'k})^2.
\label{e_fgsums}
\end{equation}
Such sums occur for the neutrons ($\rho=\nu$) as well as the protons ($\rho=\pi$).
The summation indices run over the orbitals below the shell closure (unprimed $k$)
and over the orbitals above the shell closure (primed $k'$),
and the sums obviously depend on the orbitals included in the model space. 

Let us first assume, as it happens in light nuclei (e.g., $^{16}$O or $^{40}$Ca),
that the shell closures
coincide with those of the harmonic oscillator,
in which case $j$ runs over all orbitals in a major oscillator shell, say $N$,
$j=\frac12,\frac32,\dots,N+\frac12$,
while $j'$ runs over all orbitals in the next oscillator shell $N+1$.
The first sum in Eq.~(\ref{e_fgsums}) corresponds to
\begin{equation}
\sum_{jj'}(-)^{\ell+j-1/2}(2j+1)(2j'+1)
\left(\begin{array}{ccc}
j'&j&J\\\frac12&\frac12&-1
\end{array}\right)
\left(\begin{array}{ccc}
j'&j&J\\\frac12&-\frac12&0
\end{array}\right),
\label{e_fgsum1}
\end{equation}
and can be rewritten as follows:
\begin{eqnarray}
&&\sum_j(-)^{\ell+j-1/2}(2j+1)
\sum_{j'=|j-J|}^{j+J}(2j'+1)
\left(\begin{array}{ccc}
j&J&j'\\\frac12&-1&\frac12
\end{array}\right)
\left(\begin{array}{ccc}
j&J&j'\\-\frac12&0&\frac12
\end{array}\right)
\label{e_fgsum2}\\&&-
\sum_{j=N-J+5/2}^{N+1/2}(-)^{\ell+j-1/2}(2j+1)
\sum_{j'=N+5/2}^{N+J+3/2}(2j'+1)
\left(\begin{array}{ccc}
j'&j&J\\\frac12&\frac12&-1
\end{array}\right)
\left(\begin{array}{ccc}
j'&j&J\\\frac12&-\frac12&0
\end{array}\right).
\nonumber
\end{eqnarray}
Since the sum over $j'$ in the first term of Eq.~(\ref{e_fgsum2}) is unrestricted,
orthogonality of the Wigner $3j$ coefficients implies that it vanishes.
The second term corrects for the over-counting in the first sum
and, for $J=3$, it equals
\begin{eqnarray}
&-4(N+1)(N+4)
\left(\begin{array}{ccc}
N+\frac72&N+\frac12&3\\\frac12&\frac12&-1
\end{array}\right)
\left(\begin{array}{ccc}
N+\frac72&N+\frac12&3\\\frac12&-\frac12&0
\end{array}\right)
\nonumber\\&+4(N+2)(N+4)
\left(\begin{array}{ccc}
N+\frac72&N+\frac32&3\\\frac12&\frac12&-1
\end{array}\right)
\left(\begin{array}{ccc}
N+\frac72&N+\frac32&3\\\frac12&-\frac12&0
\end{array}\right)
\nonumber\\&+4(N+2)(N+5)
\left(\begin{array}{ccc}
N+\frac92&N+\frac32&3\\\frac12&\frac12&-1
\end{array}\right)
\left(\begin{array}{ccc}
N+\frac92&N+\frac32&3\\\frac12&-\frac12&0
\end{array}\right),
\label{e_fgsum3}
\end{eqnarray}
of which it can be shown that it also vanishes.
It follows therefore that
\begin{equation}
\sum_{k'k}f^\rho_{k'k}g^\rho_{k'k}=0.
\label{e_fgsum4}
\end{equation}
The second and third sum in Eq.~(\ref{e_fgsums})
can be worked out in a similar fashion,
leading to
\begin{equation}
\sum_{k'k}(f_{k'k})^2=
\sum_{k'k}(g_{k'k})^2=
\frac{N(N+1)(N+2)(4N+11)}{(2N+3)(2N+5)}.
\label{e_fgsum5}
\end{equation}
The conclusion is therefore that,
in light nuclei with harmonic-oscillator shell closures for neutrons and protons,
the matrix~(\ref{e_sdih2}) is diagonal.
Furthermore, the energy of the lowest octupole excitation is given by
\begin{eqnarray}
E(3^-_{{\rm c}\rho})&=&
\Delta\epsilon_\rho-a_{1\rho}\frac{N_\rho(N_\rho+1)(N_\rho+2)(4N_\rho+11)}{2(2N_\rho+3)(2N_\rho+5)}
\nonumber\\&\approx&
\Delta\epsilon_\rho-\frac{a_{1\rho}}{2}\left(N_\rho^2+\frac{7}{4}N_\rho-\frac{1}{2}+\cdots\right),
\label{e_sdiena}
\end{eqnarray}
where $N_\rho$ is the major oscillator quantum number
associated with the hole orbitals below the shell closure
for the neutrons ($\rho=\nu$) and for the protons ($\rho=\pi$).

In heavier nuclei, due to the spin--orbit splitting,
unnatural orbitals intrude into the harmonic-oscillator shells.
If $N$ denotes again the oscillator quantum number
of the orbitals below the shell closure,
then the single-particle angular momentum
of the intruding orbital below the shell closure is $N+\frac32$
while it is $N+\frac52$ above the shell closure.
Because of parity, only one additional term is needed to each of the sums~(\ref{e_fgsums}),
corresponding to a particle--hole excitation between the intruder orbitals.
The corrections to the first, second and third sum of Eq.~(\ref{e_fgsums}) are, respectively,
\begin{eqnarray}
&&4(N+2)(N+3)
\left(\begin{array}{ccc}
N+\frac52&N+\frac32&3\\\frac12&\frac12&-1
\end{array}\right)
\left(\begin{array}{ccc}
N+\frac52&N+\frac32&3\\\frac12&-\frac12&0
\end{array}\right),
\nonumber\\
&&4(N+2)(N+3)
\left(\begin{array}{ccc}
N+\frac52&N+\frac32&3\\\frac12&\frac12&-1
\end{array}\right)
\left(\begin{array}{ccc}
N+\frac52&N+\frac32&3\\\frac12&\frac12&-1
\end{array}\right),
\nonumber\\
&&4(N+2)(N+3)
\left(\begin{array}{ccc}
N+\frac52&N+\frac32&3\\\frac12&-\frac12&0
\end{array}\right)
\left(\begin{array}{ccc}
N+\frac52&N+\frac32&3\\\frac12&-\frac12&0
\end{array}\right).
\label{e_fgsum6}
\end{eqnarray}
Insertion of the expressions for the Wigner $3j$ coefficients leads to the following results:
\begin{eqnarray}
\sum_{k'k}f_{k'k}g_{k'k}&=&
\frac{\sqrt{3}(N+2)(N+3)(N^2+5N+4)}{2(2N+3)(2N+5)(2N+7)},
\label{e_fgsum7}\\
\sum_{k'k}(f_{k'k})^2&=&
\frac{(N-1)N(N+1)(4N+7)}{(2N+1)(2N+3)}+
\frac{(N+1)(N+2)(N+3)(N+4)}{4(2N+3)(2N+5)(2N+7)},
\nonumber\\
\sum_{k'k}(g_{k'k})^2&=&
\frac{(N-1)N(N+1)(4N+7)}{(2N+1)(2N+3)}+
\frac{3(N+1)(N+2)(N+3)(N+4)}{(2N+3)(2N+5)(2N+7)}.
\nonumber
\end{eqnarray}
The conclusion is therefore that,
in heavier nuclei with a spin--orbit shell closure for the $\rho$ nucleon,
the matrix~(\ref{e_sdih2}) to leading orders in $N_\rho$
can be written as
\begin{equation}
\frac{a_{1\rho}}{2}
\left[\begin{array}{ccc}
\displaystyle N_\rho^2-\frac{7}{32}N_\rho-\frac{75}{64}+\cdots&~~~&
\displaystyle \imath\sqrt{3}\left(\frac{1}{16}N_\rho+\frac{5}{32}-\cdots\right)\\
\displaystyle \imath\sqrt{3}\left(\frac{1}{16}N_\rho+\frac{5}{32}-\cdots\right)&~~~&
\displaystyle -N_\rho^2-\frac{1}{8}N_\rho+\frac{5}{16}-\cdots
\end{array}\right].
\label{e_sdih2a}
\end{equation}
This shows that the off-diagonal matrix element is small
compared to the difference between the diagonal matrix elements.
It follows that the energy of the lowest octupole excitation is approximately
\begin{equation}
E(3^-_{{\rm c}\rho})\approx
\Delta\epsilon_\rho-\frac{a_{1\rho}}{2}
\left(N_\rho^2+\frac{1}{8}N_\rho-\frac{163}{512}+
\cdots\right),
\label{e_sdiena}
\end{equation}
which gives the first few terms in a $1/N_\rho$ expansion.


\begin{thebibliography}{99}
\bibitem{Butler96}
P.~A.~Butler and W.~Nazarewicz,
Rev.\ Mod.\ Phys.\ {\bf68}, 349 (1996)

\bibitem{Gaffney13}
L.~P.~Gaffney {\it et al.},
Nature {\bf497}, 199 (2013)

\bibitem{BM75}
A.~Bohr and B.~R.~Mottelson,
{\it Nuclear Structure II. Nuclear Deformations}
(Benjamin, New York, 1975)

\bibitem{Brown00}
B.~A.~Brown,
Phys.\ Rev.\ Lett.\ {\bf85}, (2000) 5300

\bibitem{Ralet19}
D.~Ralet {\it et al.},
Phys.\ Lett.\ B {\bf797}, (2019) 134797

\bibitem{Shalit63}
A.~de-Shalit and I.~Talmi,
{\it Nuclear Shell Theory}
(Academic Press, New York, 1963)

\bibitem{Talmi93}
I.~Talmi,
{\it Simple Models of Complex Nuclei.
The Shell Model and the Interacting Boson Model}
(Harwood, Chur, 1993)

\bibitem{Pandya56}
S.~P.~Pandya,
Phys.\ Rev.\ {\bf103}, (1956) 956

\bibitem{Bell59}
J.~S.~Bell,
Nucl.\ Phys.\ {\bf12}, (1959) 117

\bibitem{Lawson80}
R.~D.~Lawson,
{\it Theory of the Nuclear Shell Model}
(Clarendon Press, Oxford, 1980)

\bibitem{BM69}
A.~Bohr and B.~R.~Mottelson,
{\it Nuclear Structure I. Single-Particle Motion}
(Benjamin, New York, 1969)

\bibitem{Brussaard77}
P.~J.~Brussaard and P.~W.~M.~Glaudemans,
{\it Shell-Model Applications in Nuclear Spectroscopy}
(North-Holland, Amsterdam, 1977)

\bibitem{Rejmund99}
M.~Rejmund, M.~Schramm and K.~H.~Maier,
Phys.\ Rev.\ C {\bf59}, (1999) 2520

\bibitem{Wrzesinski01}
J.~Wrzesi\'nski {\it et al.},
Eur.\ Phys.\ J.\ A {\bf10}, (2001) 259

\bibitem{Rejmund20un}
M.~Rejmund, private communication

\end{thebibliography}
\end{document}